# Accelerated recrystallization of nanocrystalline films as a manifestation of the inner size effect of the diffusion coefficient

*S. Petrushenko*[1,2,*], *S. Dukarov*[1], *M. Fijalkowski*[2], *V. Sukhov*[1]

[1] V.N. Karazin Kharkiv National University, Kharkiv, Ukraine

[2] Technical University of Liberec Studentska, Liberec, Czech Republic

[*] Serhii.Petrushenko@tul.cz

*The work is devoted to the study of recrystallization occurring during short-term annealing of 100 nm thick polycrystalline films of copper and silver. It is found that in copper films deposited by the method of thermal evaporation onto a substrate at room temperature, a bimodal crystallite size distribution with maxima at 15 and 35 nm is observed. The bimodal distribution in copper films is preserved during annealing, which leads to a shift of both peaks of the crystallite size distribution histograms to the larger sizes region. In contrast to Cu, even in as-deposited Ag films, besides the nanosized fraction, micron-sized crystallites are present. Apparently, these grains are formed due to the phenomenon of self-annealing and weakly evolve during heating owing to grain growth stagnation. The nanosized fraction in as-deposited Ag films is represented by crystallites with the most probable size of 25 nm, which increases to 50 nm as a result of short-term annealing at the temperature of 250°C. The grain-boundary diffusion coefficient was determined, which is more than $10^{-18}$ $m^2/s$ for both films of metals. The obtained value indicates a multiple intensification of self-diffusion processes in films, the thickness of which allows us to refer them to macroscopic sample*

*Introduction*

Thin metal layers with high electrical conductivity are an important object of modern technologies. At the same time, polycrystalline films are widely used in applied, the behavior of which is significantly influenced by grain boundaries. For example, excess energy is connected with grain boundaries [1], which initiates de-wetting of films when they are heated [2, 3]. Also, important for applications is the scattering of charge carriers occurring at grain boundaries [4]. Grain boundaries under certain conditions can affect the morphology and even the phase composition of functional structures [5, 6, 7]. The energy typical of grain boundaries becomes a component of the summary free energy of the polycrystalline structure. The presence of such a summand is in many respects analogous to the surface component of the free energy, which is used to describe thermodynamic size effects, for example, the size effect of melting temperature. Thus, it can be expected that many of the size effects observed in disparate particles or thin films will also occur in nanocrystalline objects. Such effects are due to the internal, rather than external, boundaries of the sample. This allows us to call them inner size effects, and they can be caused not only by grain boundaries but also, for example, by interphase interfaces in multilayer films. Inner size effects can be observed in, generally speaking, bulk structures and change their behavior as dramatically as it occurs in the transition from bulk to nanosized particle. In particular, the excess energy of grain boundaries and other defects of polycrystal can significantly intensify diffusion processes, which will significantly accelerate the recrystallisation of samples. For example, the work [8] presents the phenomenon of self-annealing, i.e., optimization of the microstructure of films occurring at room temperature. This process is due to the high energy of defects typical of as-deposited films. Recrystallization can be an effective way of relaxation of such defects, and their high energy makes intensive grain growth possible at low and even room temperature. Thus, in the study of the temperature dependence of resistance, it has been shown [9, 10] that during the first annealing of as-deposited films, their electrical resistance irreversibly decreases. This occurs within a few tens of seconds, indicating a rapid optimization of the microstructure of the films. Exactly such recrystallization processes are a convenient way to optimize the electrical resistivity of thin films [11]. It was shown in [12, 13] that, depending on the heating kinetics, annealing of as-deposited films can both increase the thermal stability of continuous layers and provide solid-phase de-wetting of films [14]. Also, recrystallization, which due to the microstructure of the films can occur along the selected directions [15], allows ensuring the formation of contacts of copper-copper type that do not require the use of soldering [16, 17]. Also, annealing and its accompanying changes in microstructures can be used to control the magnetic properties of multilayer systems [18]. Finally, the phenomenon of recrystallization is the basis of powder technologies, which, along with the application of modern approaches, for example, spark sintering [19], can provide the formation of unique structures. In general, it can be said that, despite the availability of fundamental theoretical models [20], to date, many issues of grain growth in polycrystals are far from being fully resolved [20]. Separate difficulties arise in the study of the recrystallization of thin-film structures. Such structures may have recrystallisation features, which will be due to both a size effect related to the film thickness and an inner size effect related to the nanocrystalline nature of the films. Taking this into account, this work is devoted to the study of the influence of annealing on the microstructure of polycrystalline Cu and Ag films obtained by vacuum condensation.

*Experimental*

Samples for studies were deposited by the vacuum condensation method under conditions of oil-free vacuum. Thin films of copper or silver were deposited on cleavages of KCl single crystals and glass substrates by thermal evaporation from molybdenum boats. The thickness of the samples was determined by the quartz resonator method. The vacuum system has an oil-free pumping system based on a rotary vane pump connected to the system via a nitrogen trap and sputter ion pumps. The samples were deposited at a residual gas pressure of $10^{-6}$ mm Hg at a deposition rate of about 2 nm/s. Two types of samples were prepared. In the first of them, films after deposition were not subjected to thermal influence. In the second series, the samples after deposition were annealed at 250 °C for 2–5 minutes. Films up to 100 nm in thickness were used for TEM studies. Samples of both types were obtained within a single sputtering act. The selected thickness still allows the samples to be investigated by TEM methods, but is already large enough for the films to remain stability under the selected thermal effect. In addition, as a rule, size effects become significant in the much smaller thicknesses of samples. Thus, the chosen thickness is already sufficient for the samples to be considered bulk in the context of size effects. In some cases, films of slightly higher thickness have been used in SEM studies for a more detailed understanding of the microstructure of the cross-section. To observe the cross-section of the films, the cleavage method was used, in which thin cover glasses were used as a substrate. Upon completion of condensation, the samples were cooled to room temperature, removed from the vacuum chamber, and examined by methods of TEM and FESEM microscopy. TEM images of films of both series were obtained using SELMI EM-125 in dark and bright field modes. SELMI EMV 100BR was used for obtaining electron diffraction patterns, providing obtaining diffraction patterns in a wide beam and virtually from the entire electron-microscopic grid. To ensure the necessary sampling and to eliminate orientation factors when taking dark-field images, a series of images corresponding to different positions of the objective diaphragm on the diffraction lines were obtained. More than 2000 grains in a sample of each series were measured to construct histograms of crystallite size distribution. A Zeiss ULTRA Plus SEM microscope was used to study the morphology of the films. The radius of a circle having an area coinciding with the area of a given grain was taken as a characteristic grain size.

*Results and discussion*

Fig. 1 shows TEM images of as-deposited Cu and Ag films. As can be seen, the films of both types are highly dispersed and have a developed microstructure. At the same time, signs of labyrinth structure are observed in the Ag samples. This structure indicates a lower thermal stability of silver films compared to Cu films. In addition, silver films have a clearly pronounced bimodal microstructure: apart from nanometer crystallites even in as-deposited films, rather large formations are observed, apparently being highly defective single crystals.

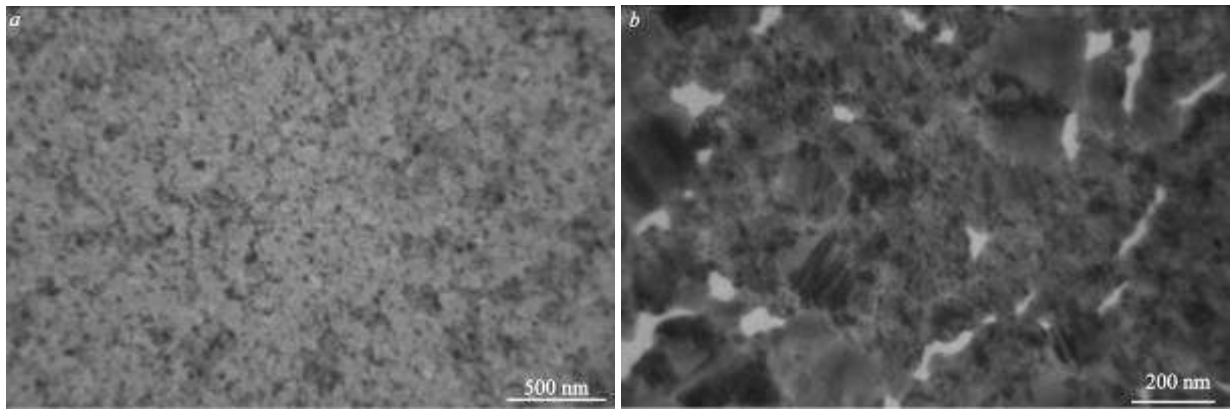

Fig. 1. TEM images of as-deposited copper (a) and silver (b) films.

The presence of two types of grains in as-deposited silver films is also observed in FESEM studies (Fig. 2). It can be seen that the crystallites observed in the film are visually divided into two classes: high-contrast crystallites of smaller size with sufficiently clear boundaries and larger but weakly contrasting grains.

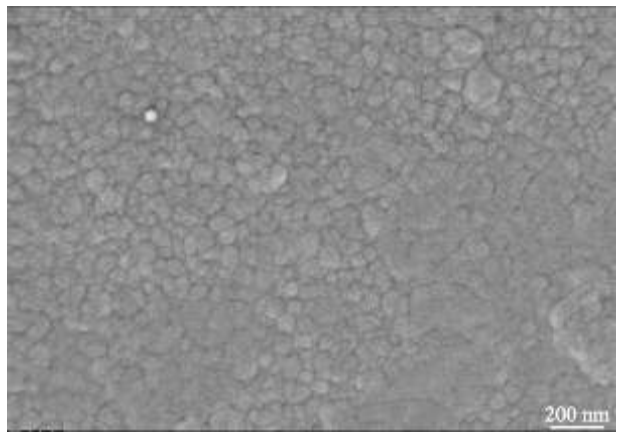

Fig. 2. FESEM image of the as-deposited Ag film.

FESEM images of as-deposited copper films are shown in Fig. 3. It can be seen that the polycrystalline nature of the films is clearly observed in the top view images. However, there are no features that would allow dividing crystallites into classes. At the same time, when using the extremely small value of the working distance, at which the maximum resolution and minimum depth of focus are realized (Fig. 3b), an indication of the complex structure of some copper grains is observed: additional elements of microstructure can be seen inside the grains. Such elements are a few nanometres in size and are usually discovered at the grain boundary.

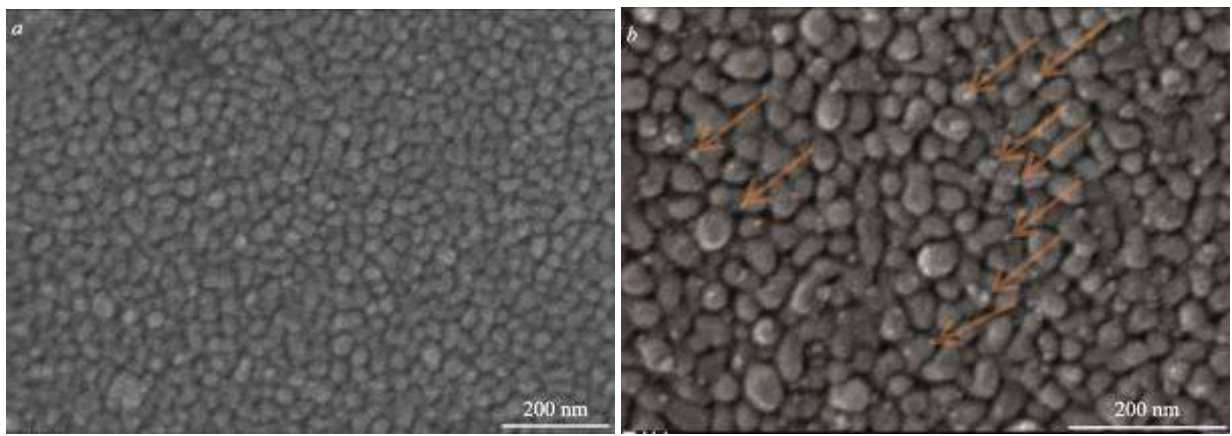

Fig. 3. FESEM images of as-deposited Cu films. Fig. 3b is obtained at the minimum WD value, optimal for observing the complex grains structure.

The microstructure of Ag films is presented in more detail in Fig. 4. As can be seen, the relatively homogeneous regions observed in the overview images have a complex internal structure with many observable boundaries. However, as TEM images obtained in the dark-field mode (Fig. 4b) show, such micron-sized regions (Figs. 1b, 2) should be attributed to single-crystal structures. Such highly defective single crystals are micron-sized, which is more typical for films of fusible metals [1]. It can be assumed that the formation of such regions is a sign of the self-annealing of films [8]. The driving force of this process is the excess energy of the as-deposited films, connected with their initially highly defective structure. The electron diffraction patterns of Cu films presented in Fig. 5 show that Ag films are characterized by broad diffraction lines. The FWHM of the lines corresponding to the (111) and (200) planes is 0.15-0.16 Å, which is very large even considering the nanocrystalline structure of the films. According to the Williamson-Hall method, the large FWHM of the diffraction peaks indicates a high level of microstresses. It is such microstresses that stimulate self-annealing of silver films.

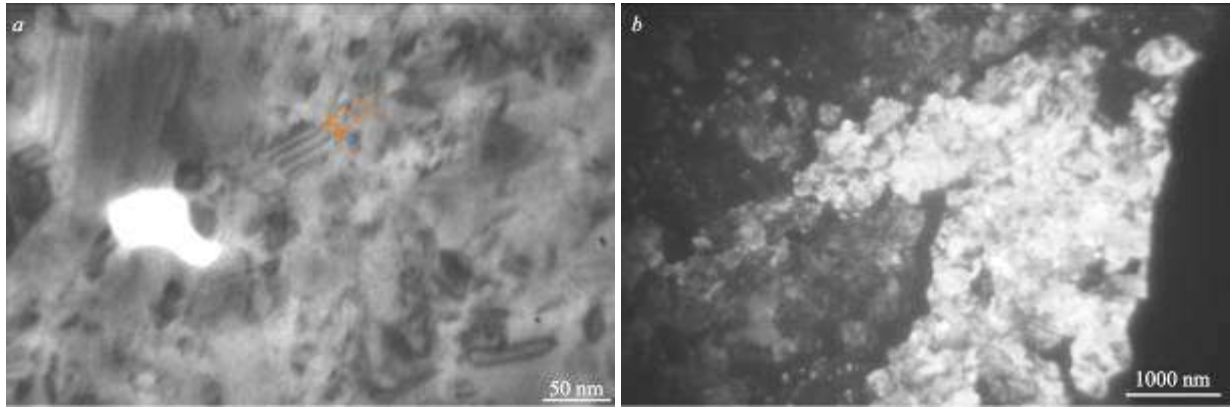

Fig. 4. Bright field (a) and dark-field (b) TEM images of the as-deposited Ag film. Arrows show twinning planes, which are one of the defects of as-deposited silver films.

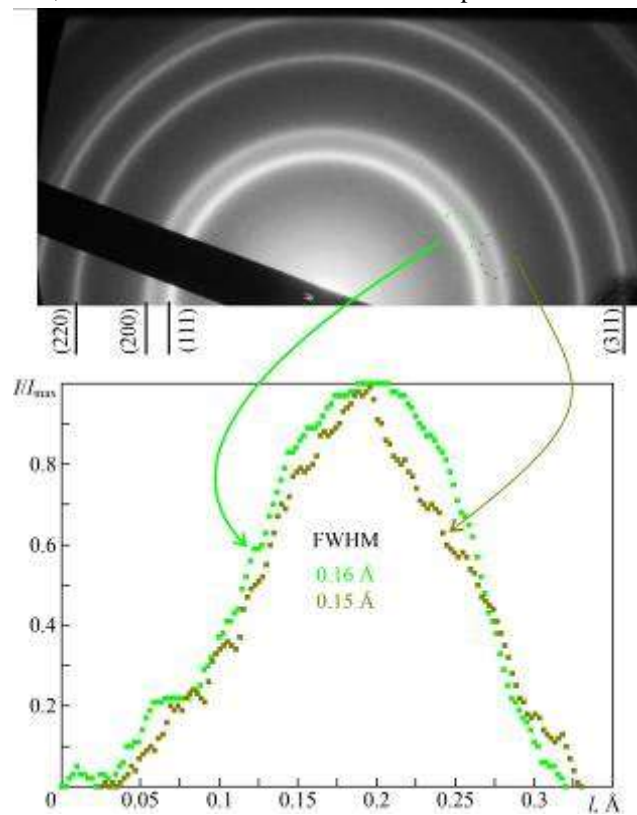

Fig. 5. The electron diffraction pattern of the as-deposited silver film and its photometric curve. The vertical axis of the photometric curve shows the intensity $(I)$ normalised to the maximum intensity on the line ($I_{max}$). The horizontal axis shows the shift ($l$) of the interplanar distance calculated from the position of the given point for which the relative brightness is calculated. A conditional point near the diffraction line is chosen as the beginning of the record

The broad diffraction lines on the electron diffraction pattern indicate that even 24 hours of exposure of the sample at room temperature preserves the highly defective state of the initial film. As expected, annealing of as-deposited films naturally leads to recrystallization, i.e., an increase in the crystallite size (Figs. 6, 7). It is important to note that in addition to recrystallization, the optimization of the thin structure of some crystallites occurs in Ag films. As a result, the number of crystal grain defects such as vacancies and dislocations is reduced, and the grain itself becomes more homogeneous in dark-field images. These crystallites correspond to the larger-sized fraction (Fig. 6b), which according to our assumption is formed by self-annealing. Both of these factors (optimization of the internal structure of crystallites and recrystallization) provide a rapid and irreversible decrease in resistance occurring during the first heating of polycrystalline films [9, 10]. At the same time, the ratios between the intensities of these processes are various in different systems. Thus, in copper films during annealing, recrystallization processes predominate, leading to the formation of large but very heterogeneous crystallites (Fig. 7). Generally speaking, annealing, which intensifies diffusion processes, as a rule, contributes to the reduction of defects and obtaining a more optimal structure. However nanosized objects observed in FESEM images of as-deposited films (Fig. 3b) prevent annealing. High-contrast borders such objects that can be likened to grooves of grains [21], are highly disoriented compared to the basic material, due to which they have high boundary energy. Exactly such developed boundaries are grain growth stoppers, i.e., the physical reason for stopping recrystallization [22, 23]. The role of such stoppers in the generation of mechanical stresses is also emphasized by the authors [24], who observed nanocavities in copper films of micron thickness. These nanocavities concentrate along grain boundaries and disappear during annealing, providing the appearance of twinning. Due to twinning, the defectivity of copper layers increases, and columnar grains can look non-columnar in cross-sec images. Thus, as the size of copper grains increases, the presence of stoppers increases the mechanical stresses developing in the growing grains that lead to crystallite defectivity. It can be expected that further optimization of the microstructure of such samples will occur mainly due to the absorption of stoppers and levelling of crystallite defects, and the recrystallization rate will already be small. At the same time, a comparison of these results with the data on irreversible decrease in resistance [9, 10] indicates that it is recrystallization that is decisive in the process of electrical conductivity improvement. The internal boundaries of crystallites, in such highly dispersed structures as thin films, apparently have a weak influence on their electrical conductivity.

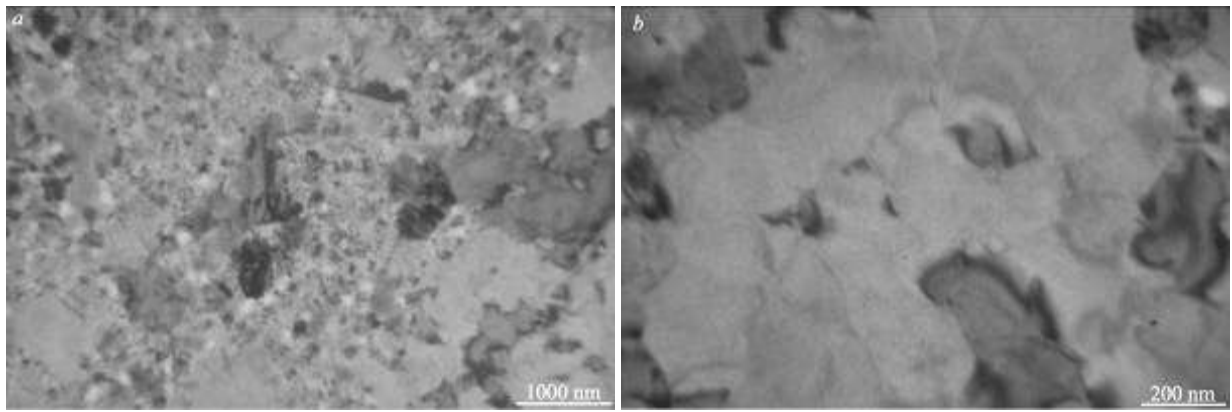

Fig. 6. TEM images of Ag films annealed immediately after condensation at 250°C for 5 min.

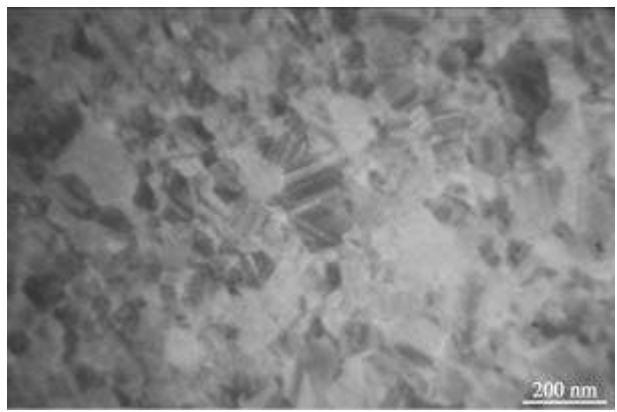

Fig. 7. TEM image of Cu-film annealed at 250°C for 5 min.

It is worth noting that the FESEM images presented in Fig. 3 are not the optimal way to measure the grain size of the nanometer range. This is primarily due to the physical nature of the contrast providing observation of individual grains in such images. Thus, it can be concluded from the analysis of the presented images that the grain spacing in Cu films is 5–10 nm. However, this is in poor agreement with the information on the electrical continuity of such samples [9, 10] and TEM images (Figs. 1, 4). Apparently, the boundaries between grains observed in Fig. 3b are not related to the absence of Cu at these lines. The finite width of such elements is related to the fact that chemical reactions occur more intensively at grain boundaries, i.e. to the increased reactive capacity of grain boundaries. Thus, it is well known that the phenomenon of carbon contamination, which is stimulated by the electron beam, is a significant problem of scanning microscopy. It is this phenomenon that makes the pictures unstable and limits the resolution of the obtained images. This problem is especially serious for images obtained at low-accelerating voltage. On the one hand, the use of low-accelerating voltage reduces the effects of carbon contamination formation. On the other hand, the low-penetration depth of beam electrons and especially the low-depth of information signal output make carbon contamination especially noticeable at low energy. Despite the fact, that the processes of carbon contamination formation are not fully understood [25], various empirical approaches have been proposed to reduce contamination [26]. In general, that the reactive activity of grain boundaries [32, 33], will stimulate the formation of carbon buildup mainly in the area of grain boundaries. Within this assumption, using FESEM images, it is possible to estimate the thickness of the grain boundary that provides the reactive activity. Considering that the boundaries in Fig. 3b belong to two grains simultaneously, such thickness can be estimated as 2-5 nm. The obtained estimate is consistent with a number of classical works that evaluate the thickness of the grain boundary [27, 28] and appears reasonable. In spite of the preliminary natures of the obtained result, the presented estimation is important for a quantitative description of the size effects that are due to the grain boundary. It is worth noting that copper films have increased reactive capacity with respect to carbon. Thus, as a rule, the image of such films deteriorates significantly after the first scanning with an electron beam. At the same time, images of silver films preserve acceptable quality even after multiple scannings of the sample. Nevertheless, in order to reduce the effects of grain-boundary contaminations, the quantitative determination of crystallite sizes was performed by dark-field images. An example of such images is shown in Fig. 8. To prevent the influence of orientation effects, several positions of the aperture diaphragm covering different regions of diffraction lines were used to obtain the series of dark-field images.

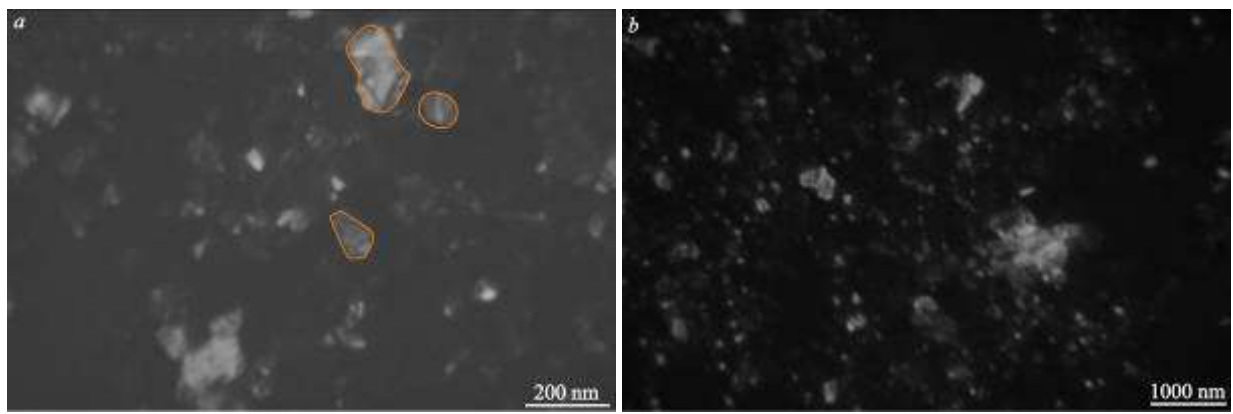

Fig. 8. Dark-field TEM images of annealed Cu (a) and Ag (b) films. The orange outline highlights copper grains having a highly defective structure after annealing. This is indicated by contrasting bands observed in the dark-field image.

Fig. 9 shows the histograms of crystallite size distribution observed in Cu and Ag films. On the vertical axis of the histograms is plotted the value $S/S_{total}$, showing the fraction of the area that crystallites of a given size have in the total area of crystallites observed in the images of this series. Since the main focus of this work is on recrystallization processes during films annealing, single micron-sized crystallites are not included in the consideration. The use of histograms of particle distribution by area, rather than by number, allows us to focus on the crystallites where the main substance of the film is concentrated. Thus, in our case, the average square deviation of the particle size is 10 nm in as-deposited and 30 nm in annealed films. Such rather large values are characteristic of vacuum condensates and are associated with the appearance of many small particles, which, however, contain an insignificant amount of substance. The transition to area distribution allows concentrating attention on those elements where the main substance of the film is concentrated. At the same time, it should be noted that the appearance of big grains in silver films is explained by self-annealing, due to which their size becomes sufficient for grain growth stagnation. Such grains preserve their size with the thermal exposure applied. They are represented by not-filled dots in the histogram. They are not taken into account in determining the most probable size of nanometer grains. Quantitative comparison of grain sizes in as-deposited metal films with those presented in the literature [29, 30, 31] is very difficult. This is due both to the different methods used by the authors to obtain the films and to some ambiguity of the grain size concept itself, which is defined somewhat ambiguously by different authors. Nevertheless, the grain size in our as-deposited films is close to that obtained by the authors using low-energy deposition methods [31].

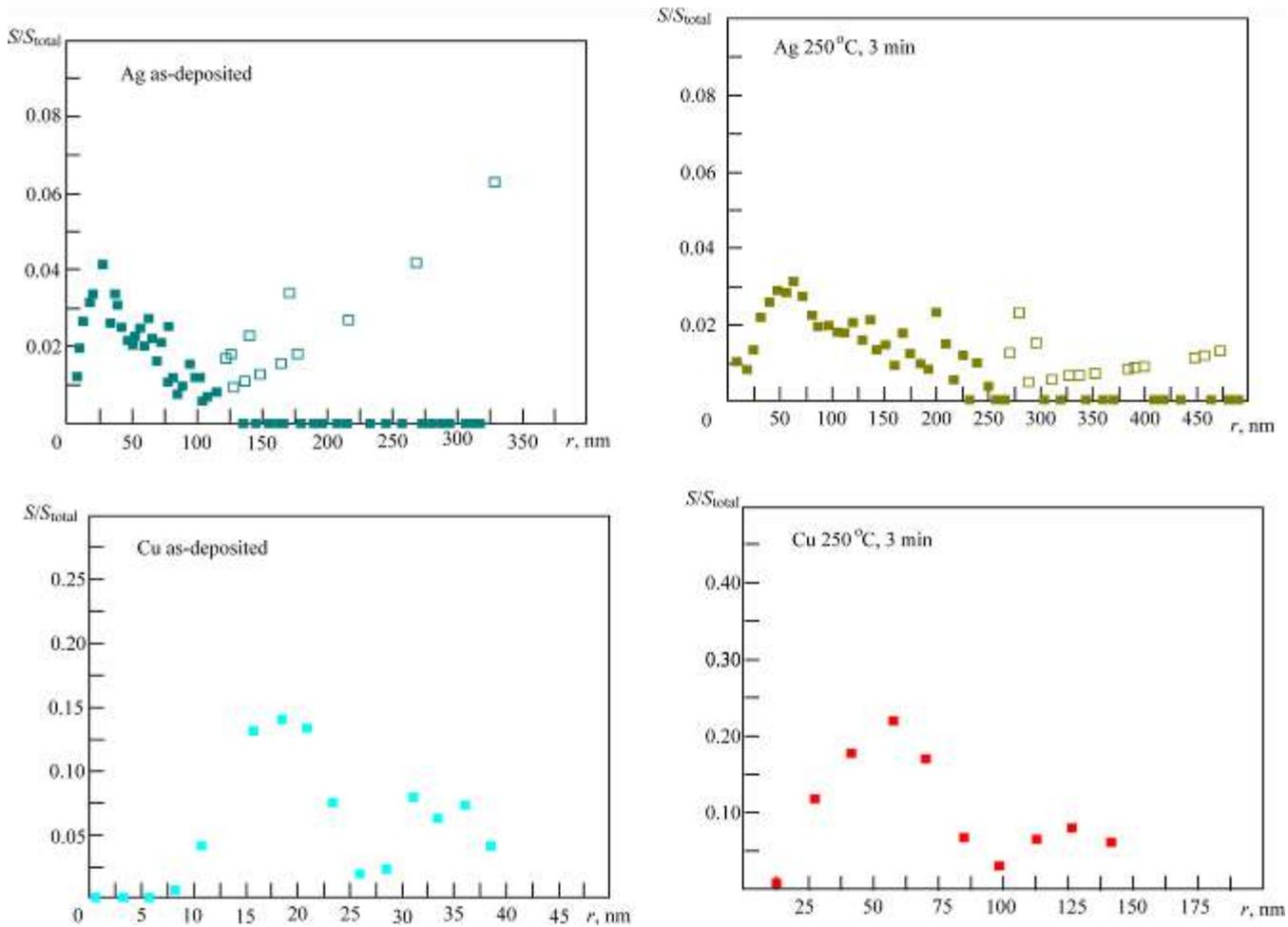

Fig. 9. Histograms of crystallite size distribution observed in as-deposited and annealed Ag and Cu films.

Diffraction studies can also provide some insight into the evolution of the crystallite sizes of the films during annealing. Fig. 10 shows a series of electron diffraction patterns obtained from copper films at different stages of annealing. When obtaining electron diffraction patterns, the method of standard was applied, in which one of the films is annealed and the second film is on the same axis with it, but located outside the heater. Generally speaking, such a method is convenient for determining the thin effects of the temperature dependence of the lattice parameter and is used, in particular, to study solubility or thermal expansion in thin films. However, in our context, it is important that both the heated film and the reference are under conditions of close energy influence of the electron beam. Thus, observation of the reference allows us to check the effect of radiation-stimulated recrystallisation, which can be caused by the electron beam. It can be seen from Fig. 10 that the diffraction lines, which are obtained actually from the whole electron-microscopic grid, are continuous, which is characteristic of nanocrystalline samples. During heating, the FWHM of the lines belonging to the heated film decreases, which is expected with increasing grain size (Fig. 10 a and b). This effect is observed for about 2 min after heating the sample. Subsequently, even after holding and multiple thermal cyclings (Fig. 10 c, d), during which the time of the sample in the heated state reaches 15–20 minutes, the FWHM of the diffraction lines practically does not change.

This indicates that two-minute heating fully completes the first, most intensive stage of recrystallisation. The same conclusion can be reached by the data of resistivity studies [9, 10], according to which the sharp decrease in film resistance occurring during its first annealing requires no more than 2 minutes. Thus, it can be stated that the first, most intensive stage of recrystallisation, which is actually studied in this work, occurs no more than 2 minutes after heating. Further, due to the stagnation of grains, recrystallisation will occur much slower with rates typical for diffusion in massive bodies. It is also worth noting that the FWHM of the diffraction lines of the reference does not practically change during prolonged exposure to the electron beam (Fig. 10). Thus, we can speak about the absence of radiation-stimulated recrystallisation in the samples, which can potentially be caused by the electron beam.

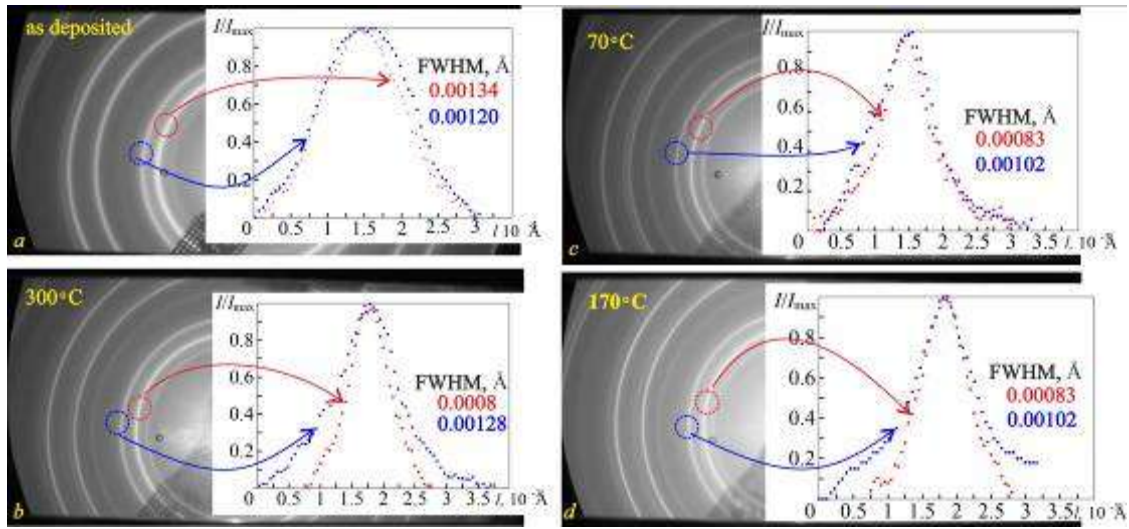

Fig. 10. Electron diffraction patterns of copper films obtained using the reference method. The complex system of lines is caused by the superposition of reflections from the crystallographic planes of the heated sample and the reference, which partially overlap. The presented photometric curves correspond to the reflection from the plane (200) of copper.

The electron diffraction pattern of the as-deposited copper film obtained without a reference is shown in Fig. 10. Apart from diffraction reflections from crystallographic planes of copper, the electron diffraction pattern shows only a diffuse halo from the film of amorphous carbon, which was used as a substrate. No oxide phases or other signs of contamination are observed. Thus, considering that the sample, the electron diffraction pattern of which is presented in Fig. 11, was transported through the atmosphere, it can be considered that the impurities do not significantly affect the recrystallisation processes occurring in the same vacuum cycle in which the samples were obtained.

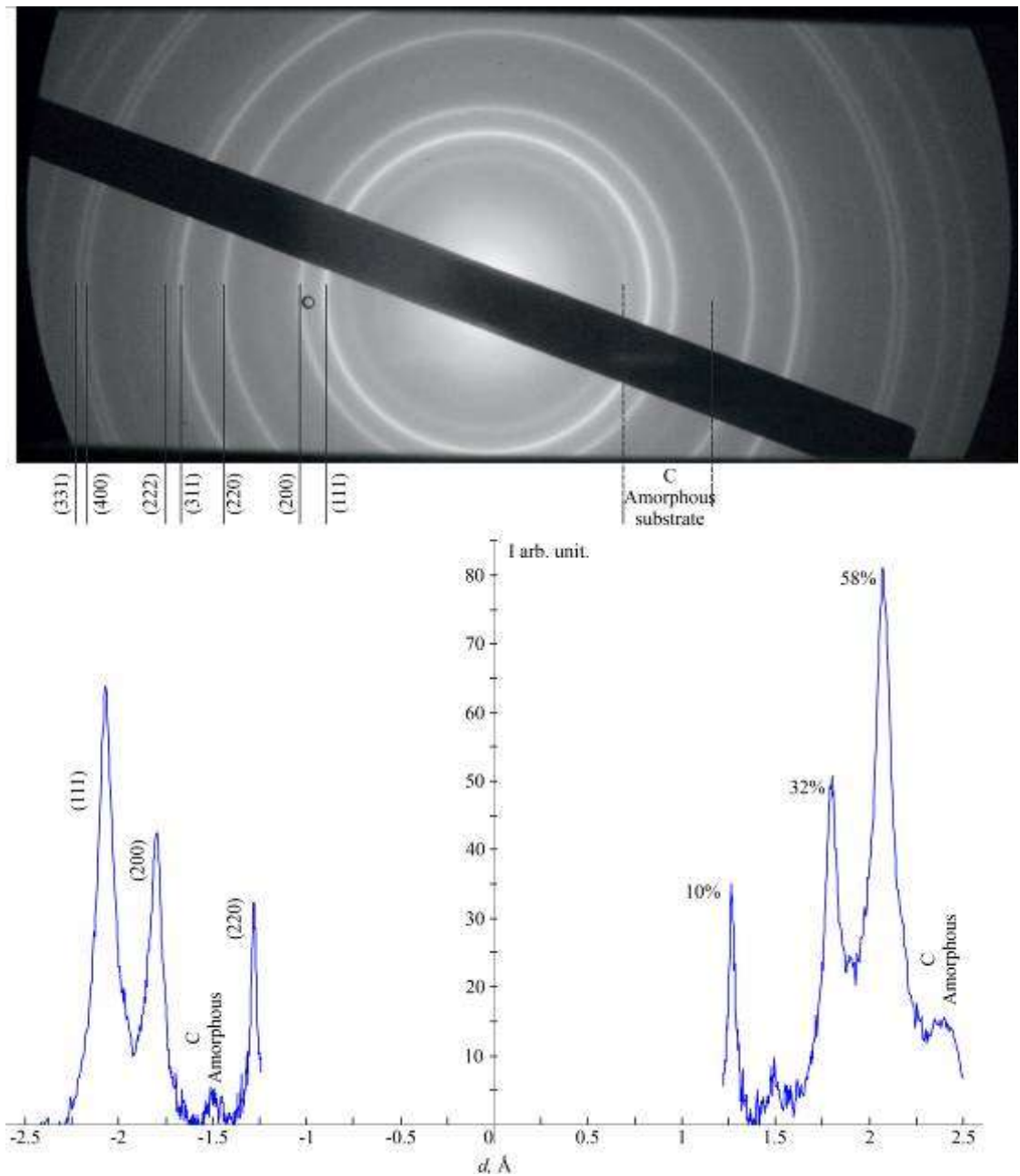

Fig. 11: The electron diffraction pattern of as-deposited copper film and the photometric plot of its central part obtained after background subtraction. The relative brightness of the lines indicated in the right part of the graph is obtained by integrating the sections of the photometric curve corresponding to the diffraction peaks.

According to classical concepts, recrystallization observed in thin films is divided into primary (normal) and anomalous (secondary) recrystallization. A characteristic feature of primary recrystallisation is that histograms of crystallite size distribution both before and after recrystallisation have a normal or lognormal type. At the same time, during secondary recrystallization, the histograms of grain size distribution initially have a bimodal type. Subsequent recrystallization of such structures is accompanied by the evolution of the type of histograms: the position of the maximum corresponding to the crystallite of smaller size remains unchanged. At the same time, its height decreases. At the same time, the maximum corresponding to larger crystallites shifts to the region of larger sizes. I.e. at secondary recrystallization, there is the absorption of dispersed fraction by grains of larger size [31]. At the same time, as can be seen from Fig. 9, the bimodal type of crystallite distribution in copper films is preserved during the performed annealing. This allows us to conclude that the observed recrystallization does not correspond to classical models and needs some explanation. Thus, the work [32] shows that columnar and non-columnar grains evolve differently during recrystallization.

At the same time, as can be seen from the images of films slices (Fig. 12), the microstructure of copper layers is close to the columnar one. This is consistent with the results of [30] according to which it is the columnar structure that is characteristic of low substrate temperatures during deposition. Due to the fact that copper belongs to the cubic syngony, the columnar structure has little effect on the electron diffraction patterns (Figs. 10, 11). Thus, the distribution of integral intensities between lines (111), (200) and (220) is 58, 32 and 10%. Similar values, which can be calculated from ICSD 64699 data (PDF 01-085-1326), are 60, 28 and 12 % for planes (111), (200) and (220), respectively. However, the columnar structure is observed morphologically (Fig. 12). Thus, crystallites formed on the substrate during condensation have practically no distinguished crystallographic direction, but after their formation they grow in the vertical direction, i.e. in the direction of the main flow of substance.

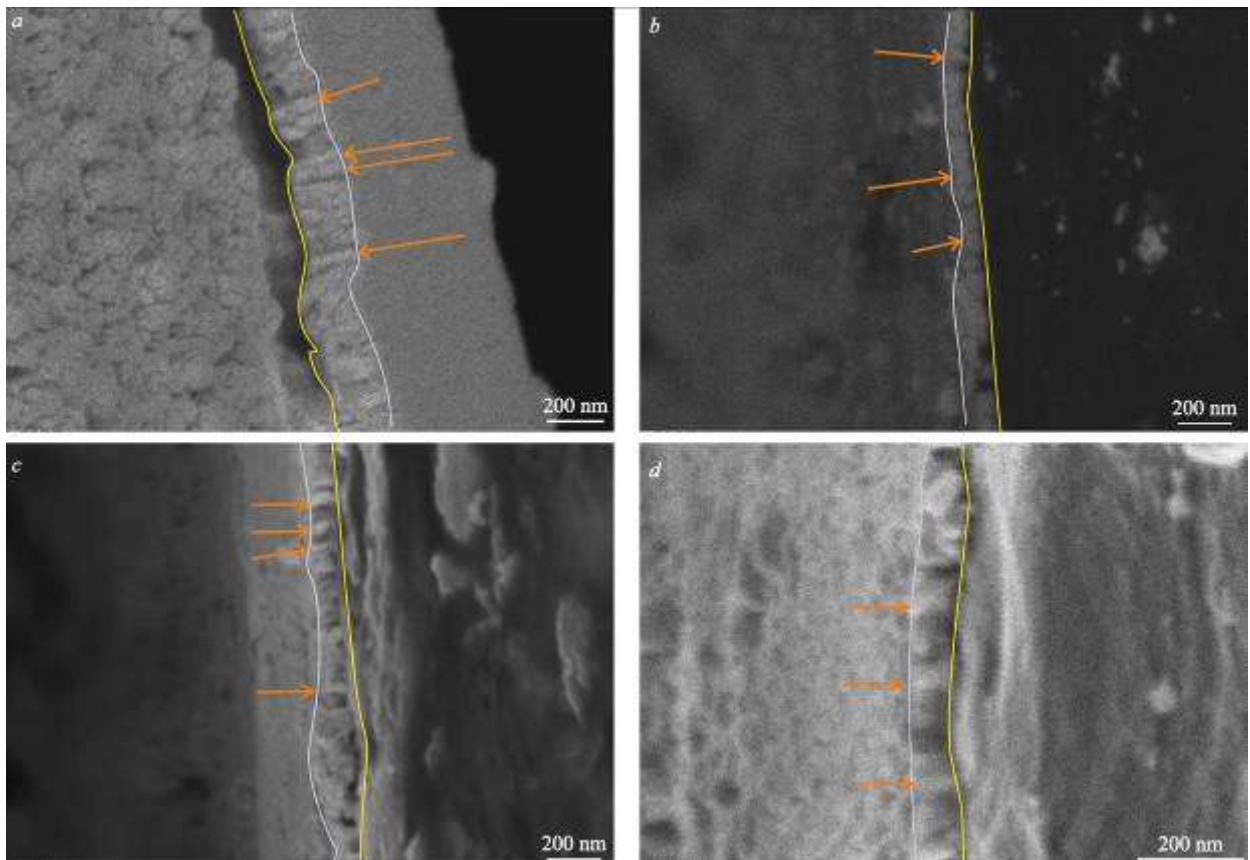

Figure 12: A series of chips and slices showing the microstructure of the cross-section of copper films. Image (b) taken using by ESB detector. Film boundaries are visualised by coloured lines. The arrows indicate the places of confident observation of columnar grains.

The columnar nature of the films affects the kinetics of their recrystallisation. Thus, due to this structure, recrystallisation is provided by the side surfaces of the grain, while the upper and lower sides of the it do not participate in the process. In addition, such a structure turns out to be more convenient for grain sizes determination, which is especially important when using the SEM technique. Nevertheless, even with a non-columnar structure, the use of dark-field TEM microscopy allows us to determine the sizes of crystallographically disoriented grains quite correctly, even if they are overlapped on 2D projections of each other. The bimodal type of particle size distribution diagrams requires separate consideration (Fig. 9). As can be seen in Fig. 13 copper films have significant roughness, which is due to the presence of some number of mushroom-shaped grains, the height of which exceeds the film thickness by 45–50 percent (Fig. 14).

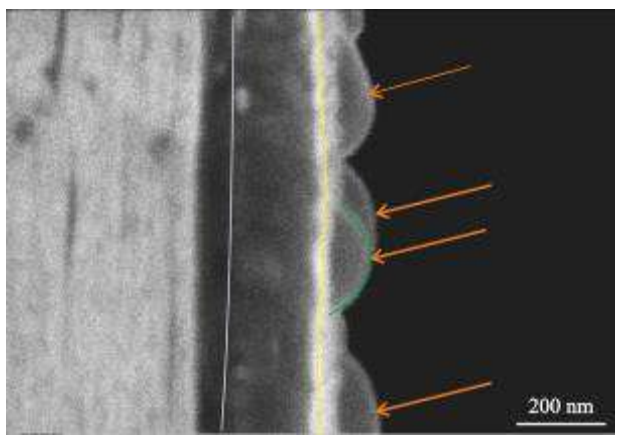

Fig. 13. A slice image of an as-deposited copper film showing the rough structure of the sample.

It can be seen (Fig. 14) that such mushroom-shaped elements have a complex structure (Fig. 12c) and a small area of contact with the basic film (Fig. 14b). That is, the inflow of substance into such elements of the morphological structure from the side of the basic film will be limited. It is reasonable to expect that due to the restriction of substance flow, the recrystallization of grains constituting such elements of microstructure will occur differently than the recrystallization of the main array of grains. The influx of substance into crystallites that grow in mushroom-shaped elements will be mainly due to those grains that are in this morphological element, and the rest of the film substance will be inaccessible to them. Due to this, it is possible to speak about grains of two types, which form two peaks on the distribution histograms. Due to differences in the availability of substance for the growth of such grains, their evolution will occur independently of each other. Thus, the recrystallization of copper films can be described as a process of primary recrystallization, which occurs independently with grains of two types. One of them provides the growth of grains of the main bulk of the film. Such grains have free access to the surrounding substance. The second, which corresponds to the more dispersed fraction, corresponds to the grains that make up the mushroom-shaped elements. For such grains, which according to FESEM microscopy data belong to the dispersed fraction, the substance of the surrounding film is limitedly available. This imposes restrictions on their growth kinetics, ensuring the preservation of the bimodal type of distribution histograms.

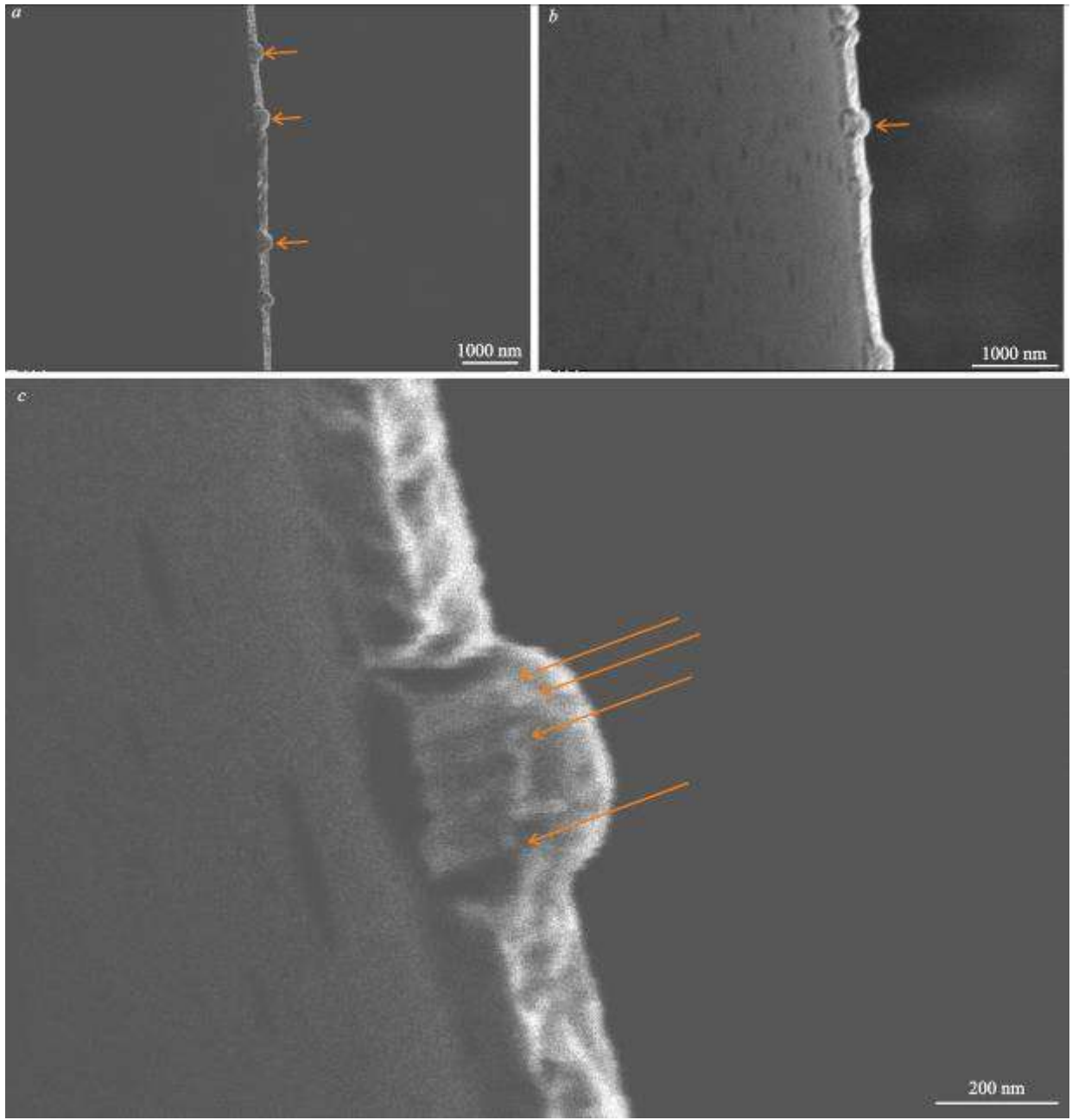

Fig. 14. FESEM images of mushroom-shaped elements in copper films. The images show the internal structure of mushroom-shaped elements (c) and the weak contact of such structures with the core film material (b).

In contrast to copper films, histograms of crystallite distribution in silver films indicate that the recrystallization of these samples (at least for the highly dispersed fraction) has a primary nature. Upon annealing, the peak of the histogram shifts to the region of larger sizes, and there is no change in the histogram appearance. At the same time, unlike copper films, silver layers do not have a columnar structure, but a layered structure (Fig. 15). I.e., such films consist of thin layers, overlapped one on another. It is the layered structure of the films that explains the complex appearance of TEM images of as-deposited films (Fig. 4). Such images provide integral information formed by the entire depth of the film. It is natural, that the presence of internal boundaries parallel to the substrate complicates the observed patterns. At the same time, due to the orientating effect of crystallites, the growth of the upper layers is close to epitaxial [15], due to which such grains are combined in dark-field images. Due to the layered structure of silver films, they obtain an additional mechanism of microstructure evolution - recrystallization of such samples can occur not only by horizontal but also by vertical mass transfer. This mechanism is very high-speed for thin-film structures [34]. Due to the small thickness of the layers, it is the vertical mass transfer that plays the main role at the initial stages of recrystallization, providing optimization of the thin structure of crystallites. Also, it is the layered structure and the associated excess energy that provide the possibility of self-annealing. In this context, silver films resemble micron films of copper studied in the work [33]. The authors showed that copper layers with a non-columnar structure tend to anisotropic grain growth, with the advantage of vertical direction. After rapid vertical optimization of the layers, providing optimization of the thin structure of grains, further growth of crystallites occurs, ensuring an increase in their size in the plane of the substrate.

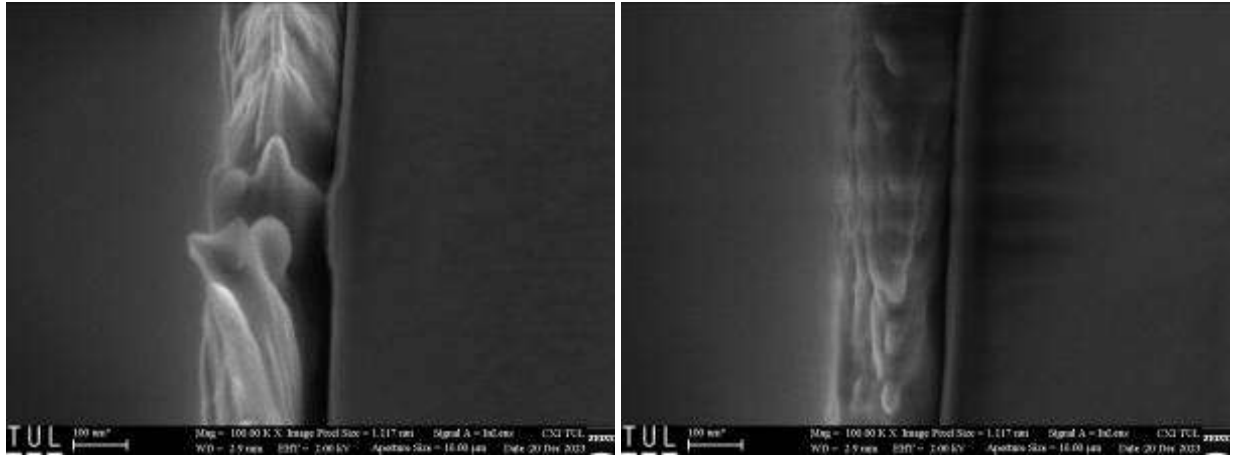

Fig 15. Slice images of Ag films showing their layered structure.

Based on the presented data, the layered structure of silver films can explain their low thermal stability [9, 10], which does not improve with increasing film thickness. The films under study are so eager to solid-phase de-wetting that they often collapse under the action of a low-voltage beam of the electron microscope (Fig. 16). In such a structure, the inner size effect is provided not by the lateral grain size, but by its thickness, which amounts to units of nanometres. Due to this, the intensity of mass transfer increases dramatically, and each of the layers can to some extent be regarded as an element of the microstructure that evolves independently.

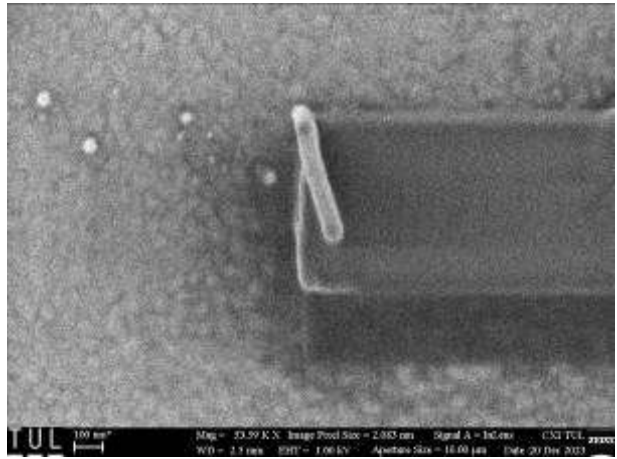

Fig. 14. FESEM image of a silver film showing the effect of the action of the electron beam. During scanning, the film peels off and rolls up into a tube.

According to classical phenomenological models describing recrystallization kinetics [31], the grain size change can be described using the equation:

$$r^m - r_0^m = at, \qquad (1)$$

where

$$a = a_0 e^{\frac{E_a}{kT}}.$$

The degree exponent in equation (1) is usually assumed to be 2, but in real experiments, it may differ from its theoretical value. The pre-exponential multiplier in expression (1) is assumed to be independent of temperature, and the value $E_a$ - represents the recrystallization activation energy. Although the movement of grain boundaries cannot be directly called a diffusion process, the form of equation (1) and general considerations indicate that the driving force of recrystallization is the self-diffusion of components. In accordance with [35], the motion of intergranular boundaries during recrystallization can be described by the introduction of some conditional grain-boundary diffusion coefficient $D_r$. To estimate it, we can [35] use an expression similar to that used to describe diffusive mass transfer:

$$D_r = \frac{x^2}{2t}, \qquad (2)$$

where $x$ is the displacement of grain boundaries, while $t$ – is the time of thermal exposure. Accepting that $x$ in our case is equal to the difference of the most probable radii of crystallites corresponding to the low-dimensional maximum of the distribution histograms, we can obtain that $D_r$ for the silver film is about $10^{-18}$ m$^2$/s. A similar value will be obtained when considering the evolution of the low-dimensional maximum of the histograms of crystallite distribution in copper films (Fig. 9). For the right maximum of the distribution histogram $D_r$, is approximately an order of magnitude larger. Some surprise is caused by the fact that the value of $D_r$ turned out to be noticeably larger for the right peak of the histogram, corresponding to larger grain sizes. At the same time, a typical tendency for nanosized structures is the increase of the size effect with decreasing size. The observed effect can be explained by the conditions of limited geometry, in which grains of small-sized fraction, forming mushroom-shaped particles, are located. The impossibility of the exchange of the substance with the surrounding film and the high density of boundaries squeezed by the outer surface are the factors reducing the recrystallization intensity. In general, it can be said that the obtained value is about 6 and 9 orders of magnitude higher than the coefficient of volume self-diffusion for silver and copper films, respectively. It should be noted that the direct identification of $D_r$ with the atomic diffusion coefficient is impossible. Thus, according to [35], $D_r$, as a rule, is 4–5 orders of magnitude higher than the value of the bulk diffusion coefficient and about 100 times higher than the grain-boundary diffusion coefficient. That is, the obtained $D_r$ value indicates that the diffusion coefficient for nanosized crystallites increases by a factor of 10 and 1000–10000 for silver and copper films, respectively. It should be noted that, according to resistive studies, the irreversible decrease in the resistance of films during their heating occurs within 10–30 s [9, 10]. During subsequent heating-cooling cycles, the temperature dependence of resistance has a linear and, with sufficient thickness, reproducible nature. Since, according to theoretical models, the recrystallization rate decreases with increasing grain size (or its radius of curvature) [36, 37], we can expect that the structure that emerged with an irreversible decrease in the resistivity of the films will already be stable.

Thus, given the data from the resistive studies, it becomes clear that the microstructure of the films that correspond to the constructed distribution histograms (Fig. 9) is formed not within 200 s, but much faster. I.e., the values of the diffusion coefficient increase presented above should be considered as a lower estimate of the size effect observed in polycrystalline films with nanometer crystallite size.

It is important to note that the chosen thickness of the films is already large enough for them to be considered massive in the context of the usual size effects. For example, in the work [38], an increase in the diffusion coefficient was observed for binary films with a thickness of 5 nm. Qualitative intensification of diffusion formation of solid solutions in films with a mass thickness of only 2.5 nm was found in the work [39]. The influence of thickness on hydrogen diffusion in palladium films with thicknesses from 22 to 135 nm was studied by the authors [40]. The authors showed that the hydrogen diffusion coefficient in 22 nm of thickness palladium films decreases by 2–3 orders of magnitude compared to the bulk state. In films of 46.3 nm in thickness, the reduction of the diffusion coefficient in comparison with the bulk state is only about one order of magnitude.

Thus, taking into account the thickness of the films chosen for the study, it can be stated that one of the manifestations of the inner size effect is observed in the work. This class of size effects is due to the fact that a bulk sample has many internal boundaries, with which some excess energy can be associated. In our case, such surfaces are grain boundaries and, to a lesser extent, boundaries of crystalline defects. This additional energy leads to a change in the free energy, which is a factor determining the intensity of diffusion processes [41, 42, 43]. Thus, the appearance of an additional summand in the free energy of the system stimulates processes that allow its optimization. In the case of heating samples to pre-melting temperatures, this factor leads to a decrease in the melting temperature of polycrystalline [1] and multilayer [44] films. At low-temperature annealing, the surface summand, connected with the internal nanostructure, can provide a decrease in the activation energy of diffusion. It is this change that ensures the size effect of diffusion in bulk nanocrystalline metal films.

*Conclusions*

The kinetics of recrystallization of Cu and Ag films, which occurs during short-term annealing of polycrystalline films with nanometer grain size, has been studied. It is shown that the bimodal type of the size distribution of crystallites, constituting Cu films, remains after their annealing. Annealing naturally leads to the enlargement of the microstructure of films of both metals. The grain-boundary diffusion coefficient determined for the investigated films is about $10^{-17}$–$10^{-18}$ m$^2$/s, which is 6 and 9 orders of magnitude higher than the value of the bulk diffusion coefficient for silver and copper films, respectively. The obtained value indicates a significant intensification of self-diffusion in nanocrystalline films, due to which the diffusion coefficient increases by a factor of 10-10000 compared to the tabular value. Due to the large thickness of the studied films, the observed increase in the diffusion coefficient should be attributed to inner size effects, which can be explained by the energy of grain boundaries.

*Acknowledgment*

The work was supported by the Ministry of Education and Science of Ukraine and within the MSCA4Ukraine project funded by the European Union. Views and opinions expressed are however, the those of the author(s) only and do not necessarily reflect the views of the European Union. Neither the European Union, nor the MSCA4Ukraine Consortium as a whole, nor individual member institutions of the MSCA4Ukraine Consortium can be held responsible for them.